\documentclass[twocolumn,superscriptaddress,nofootinbib,preprintnumbers,showpacs,tightenlines,notitlepage]{revtex4}
\usepackage[latin1]{inputenc}
\usepackage{graphicx}
\usepackage{amssymb}
\usepackage{color}
\usepackage{float}
\usepackage{amsmath}
\usepackage{amsfonts}
\usepackage{dcolumn}
\usepackage{hyperref}
\usepackage{amsthm}
\usepackage{color}
\usepackage{bm}

\def\uudot{\dot{u}}
\def\3nab{\tilde{\nabla}}

\def\la {\langle}
\def\ra {\rangle}

\def\be {\begin{equation}}
\def\ee {\end{equation}}
\def\ba {\begin{eqnarray}}
\def\ea {\end{eqnarray}}

\newcommand{\bra}[1]{\left(#1\right)}

\newcommand{\brac}[1]{\left\{#1\right\}}

\newcommand{\sfr}[2]{{\textstyle\frac{#1}{#2}}}

\newcommand{\lc}{\varepsilon}
\newcommand{\lb}{\{}
\newcommand{\rb}{\}}
\newcommand{\E}{{\mathcal E}}
\renewcommand{\H}{{\mathcal H}}
\newcommand{\barray}{\begin{array}}
\newcommand{\earray}{\end{array}}
\newcommand{\e}{e}
\newcommand{\N}{N}

 \newcommand{\nab}{\nabla}

\newcommand \ep {\epsilon}

\newcommand \om {\omega}

\newcommand{\udot}{{\mathcal A}}
\newcommand{\hh}{{\mathcal H}}
\begin{document}

\title{Rotating and twisting locally rotationally symmetric spacetimes: a general solution}
\author{Sayuri Singh}
\email{sayurisingh22@gmail.com }
\affiliation{Astrophysics and Cosmology Research Unit, School of Mathematics, Statistics and Computer Science, University of KwaZulu-Natal, Private Bag X54001, Durban 4000, South Africa.}
\author{George F. R. Ellis}
 \email{george.ellis@uct.ac.za}
 \affiliation{Department of Mathematics and Applied Mathematics and ACGC, University of Cape Town,
Cape Town, South Africa.}
 \author{Rituparno Goswami}
\email{Goswami@ukzn.ac.za}
\affiliation{Astrophysics and Cosmology Research Unit, School of Mathematics, Statistics and Computer Science, University of KwaZulu-Natal, Private Bag X54001, Durban 4000, South Africa.}
\author{Sunil D. Maharaj}
\email{Maharaj@ukzn.ac.za}
\affiliation{Astrophysics and Cosmology Research Unit, School of Mathematics, Statistics and Computer Science, University of KwaZulu-Natal, Private Bag X54001, Durban 4000, South Africa.}

\begin{abstract}
In this paper we derive a general solution for the most general rotating and twisting locally rotationally symmetric spacetimes. This is achieved in three steps. First we decompose the manifold via 1+1+2 semi-tetrad formalism that yields a set of geometrical and thermodynamic scalars for the spacetime. We then recast the Einstein field equations in terms of evolution and propagation of these scalars. It is then shown that this class of spacetimes must possess self similarity and we use this property to solve for these scalars, thus obtaining a general solution. This solution has a number of very interesting cosmological or astrophysical consequences which we discuss in detail.
\end{abstract}
 
\maketitle

\section{Introduction}

{\em Locally Rotationally Symmetric} (LRS) spacetimes are those that possess a continuous isotropy group at each point which generally implies the existence of a multiply-transitive isometry group acting on the spacetime manifold \cite{Ellis_1967,Ellis_1968}. It is well known that isotropies around a point in any spacetime with a fluid can be a 3-dimensional or 1-dimensional subgroup of the full group of  isometries; they necessarily leave the normalised 4-velocity of the matter flow invariant. The 3-d case  implies isotropy at every point, yielding the  Friedmann-Lema\^{i}tre-Robertson-Walker (FLRW) models, while the 1-d case corresponds to anisotropic and in general spatially inhomogeneous models. These  have a preferred spacelike direction $e^a$ orthogonal to the fluid flow 4-vector $u^a$: all spatial directions orthogonal to $e^a$ and $u^a$ are geometrically identical. \\

In the case of a perfect fluid, these spacetimes are split into three classes as described in Section \ref{sec:3}, depending on whether the vorticity component $\Omega$ along the direction $e^a$ of the fluid and 2-dimensional twist $\xi$ of the vector field $e^a$ are zero or not (they cannot both be non-zero in this case). However for an imperfect fluid - for example if there is  an entropy flux - both can be non-zero. In a previous paper \cite{Singh:2016qmr} we obtained a set of field equations and integrability conditions for the imperfect fluid case. We also proved that the LRS spacetimes with nonzero rotation and spatial twist must be self similar. In this paper, we extend that work by obtaining a general solution to the field equations for this situation. This is achieved by using the property of the self similarity. Also we show that we may specify an equation of state for the isotropic pressure at an initial Cauchy surface for particular applications.\\

In Section \ref{Sec2}, the semi-tetrad formalism used is introduced in a general form. In Section \ref{sec:3} it is restricted to the case of LRS fluid spacetimes. In Section \ref{Sec4}, a reduced set of field equations is obtained with self-similar variables. In section \ref{Sec5}, the general solution to the field equations is obtained for the case of LRS fluids with non-zero    rotation and spatial twist. In Section \ref{Sec6}, their properties are discussed in both cosmological and astrophysical scenarios.

\section{LRS spacetimes in semi-tetrad formalism}\label{Sec2}
Due to the symmetries of LRS spacetimes, a 1+1+2 semi tetrad covariant formalism (which is a natural extension of local 1+3 decomposition), is well suited for describing the geometry, as in this formalism the field equations become a set of coupled differential equations in covariantly defined scalar variables. In the next two subsections we briefly discuss the 1+3 and  semitetrad formalisms, and in the third  subsection we present the field equations in terms of covariantly defined geometrical scalar variables.

\subsection{1+3 decomposition of spacetime}
The 1+3 decomposition provides a covariant description of the spacetime in terms of 3-vectors, scalars and projected symmetric trace-free (PSTF) 3-tensors \cite{EllvanEll}. This is helpful for understanding various physical and geometrical aspects of relativistic fluid flows. With respect to a timelike congruence, the spacetime can be locally decomposed into time and space parts. Such a timelike congruence can be defined by the matter flow lines, with the \textit{four-velocity} defined as
\be
u^a = \frac{dx^a}{d\tau}, \quad \mbox{with} \quad u^a u_a = -1 ,
\ee
where $\tau$ is proper time along the flow lines. Given the four-velocity $u^a$, we have unique parallel and orthogonal \textit{projection tensors}
\ba
U^a{}_b &=& -u^a u_b ,
\\
h^a{}_b &=& g^a{}_b+u^au_b,
\ea
where $h^a{}_b$ is the projection tensor that projects any 4-d vector or tensor onto the local 3-space orthogonal to $u^a$ which has volume element $\epsilon_{abc}:=\eta_{abcd} u^d.$\\

From this, it follows that
we have two well defined directional derivatives. The vector $u^{a}$ is used to define the \textit{covariant time derivative} along the flow lines (denoted by a dot) for any tensor 
$ S^{a..b}{}_{c..d}$, given by 
\be
\dot{S}^{a..b}{}_{c..d}{} = u^{e} \nab_{e} {S}^{a..b}{}_{c..d}\;.
\ee
The tensor $h_{ab}$ is used to define the fully orthogonally \textit{projected covariant derivative} $D$ for any tensor $
S^{a..b}{}_{c..d} $: 
\be D_{e}S^{a..b}{}_{c..d}{} = h^a{}_f
h^p{}_c...h^b{}_g h^q{}_d h^r{}_e \nab_{r} {S}^{f..g}{}_{p..q}\;,
\ee 
with total projection on all free indices. 
In this way, the covariant derivative of $u^a$ can be decomposed as
\be
\nabla_a u_b = -u_a A_b +\frac13\Theta h_{ab}+\sigma_{ab}+\ep_{abc}\om^{c}.
\ee
Here $A_b=\dot u_b$ is the acceleration, $\Theta=D_au^a$ represents the expansion of $u_a$, 
$\sigma_{ab}=\bra{h^c{}_{\left( a \right.}h^d{}_{\left. b \right)}-\sfr13 h_{ab} h^{cd}}D_cu_d$ is the shear tensor that denotes the rate of distortion and $\om^{c}$ is the vorticity vector denoting the rotation.

The Weyl tensor is split relative to $u^a$ into the \textit{electric} and \textit{magnetic Weyl curvature} parts as
\begin{eqnarray}
E_{ab} &=& C_{abcd}u^bu^d =E_{\la ab\ra}\;,
 \end{eqnarray}
and
\begin{eqnarray}
H_{ab} &=& \sfr12\ep_{ade}C^{de}{}_{bc}u^c =H_{\la ab\ra}  \;,
\end{eqnarray}
where angle brackets represent the orthogonal symmetric trace-free part (so $\sigma_{ab}=\sigma_{\la ab\ra})$.

The energy momentum tensor of matter can be decomposed similarly as 
\be
T_{ab}=\mu u_au_b+q_au_b+q_bu_a+ph_{ab}+\pi_{ab}\;,
\ee
where $p=(1/3 )h^{ab}T_{ab}$ is the isotropic pressure, $\mu=T_{ab}u^au^b$ is the energy density, $q_a=q_{\la a\ra}=-h^{c}{}_aT_{cd}u^d$ is the 3-vector that defines the heat flux, and $\pi_{ab}=\pi_{\la ab\ra}$ is the anisotropic stress.

\subsection{1+1+2 decomposition of spacetime}
The 1+1+2  decomposition is a natural extension of the 1+3 decomposition in which the 3-space is further decomposed with respect to a given spatial direction, i.e., we now have another split along a preferred spatial direction such as the case which occur in LRS models  \cite{Clarkson:2002jz,Betschart:2004uu,Clarkson:2007yp} . We choose a spacelike vector field $e^a$  such that
\be
u^a e_a =0 \quad \mbox{and} \quad e^a e_a=1.
\ee
The new projection tensor is given by
\be \label{N1}N_a^{~b}\equiv
h_a^{~b}-\e_a\e^b=g_{a}^{~b}+u_au^b-\e_a\e^b\, . \ee
This tensor projects vectors onto local 2-spaces orthogonal to both $u^a$ and $e^a$, defined as \textit{sheets}. Thus
\be\label{N2}
e^a N_{ab}=0=u^a N_{ab}, \quad N^a{}_a=2.
\ee
This spacelike vector now naturally introduce two new derivatives, which for any tensor $ \psi_{a...b}{}^{c...d}  $: 
\ba
\hat{\psi}_{a..b}{}^{c..d} &\equiv & e^{f}D_{f}\psi_{a..b}{}^{c..d}~\label{eq:hat}, 
\\
\delta_f\psi_{a..b}{}^{c..d} &\equiv & N_{a}{}^{p}...N_{b}{}^gN_{h}{}^{c}..
N_{i}{}^{d}N_f{}^jD_j\psi_{p..g}{}^{i..j}\;.\label{eq:delta}
\ea 
The derivative (\ref{eq:hat}) along the $e^a$ vector field in the surfaces orthogonal to $ u^{a}$ is called the hat-derivative, while the derivative (\ref{eq:delta}) projected onto the sheet is called the $\delta$-derivative. This projection is orthogonal to $u^a$ and $e^a$ on every free index.

In the $1+1+2$ splitting, the 4-acceleration, vorticity and shear split in this way as
\ba
\uudot^a&=&\udot \e^a+\udot^a,\\
\omega^a&=&\Omega \e^a+\Omega^a,\\
\sigma_{ab}&=&\Sigma\bra{\e_a\e_b-\sfr{1}{2}\N_{ab}}+2\Sigma_{(a}\e_{b)}+\Sigma_{ab}.
\ea
For the electric and magnetic Weyl tensors we get
\ba
E_{ab}&=&{\cal E}\bra{\e_a\e_b-\sfr{1}{2}\N_{ab}}+2{\cal E}_{(a}\e_{b)}+{\cal E}_{ab},\\
H_{ab}&=&{\cal H}\bra{\e_a\e_b-\sfr{1}{2}\N_{ab}}+2{\cal H}_{(a}\e_{b)}+{\cal
H}_{ab}.
\ea
Similarly, the fluid variables $q^a$ and $\pi_{ab}$ are split as follows
\ba
q^a&=&Q \e^a+Q^a,\\
\pi_{ab}&=&\Pi\bra{\e_a\e_b-\sfr{1}{2}\N_{ab}}+2\Pi_{(a}\e_{b)}+\Pi_{ab}.
\ea
By decomposing the covariant derivative of $e^a$ in the directions orthogonal to $u^a$ into it's irreducible parts, we get
\be 
 {D}_{a}e_{b} = e_{a}a_{b} + \frac{1}{2}\phi N_{ab} + 
\xi\epsilon_{ab} + \zeta_{ab}~, 
\ee
where 
\ba 
a_{a} &\equiv & e^{c}{\rm D}_{c}e_{a} = \hat{e}_{a}~, \\ 
\phi &\equiv & \delta_ae^a~, \\  \xi &\equiv & \frac{1}{2} 
\epsilon^{ab}\delta_{a}e_{b}~, \\ 
\zeta_{ab} &\equiv & \delta_{\lb a}e_{b \rb }~.
\ea
Here, $\epsilon_{ab}=\epsilon_{[ab]}$ is the volume element on the sheet, $\phi$ represents the \textit{spatial expansion of the sheet},  $\zeta_{ab}$ is the \textit{spatial shear}, i.e., the distortion of the sheet, $a^{a}$ its \textit{spatial acceleration} (i.e. deviation from a geodesic), and $\xi$ is its spatial \textit{vorticity}, i.e., the ``twisting'' or rotation of the sheet. 

\section{LRS spacetimes and field equations}
The basic property of fluid filled LRS spacetimes is that there exists a unique, preferred spatial direction at every point, covariantly defined, which creates a local axis of symmetry.
Hence the 1+1+2 decomposition described in the previous section is ideally suited for the study of these spacetimes as we can immediately see that if we choose the spacelike unit vector $e^a$ along the preferred spatial direction, then by symmetry all the sheet vectors and tensors vanish identically:
\ba\label{eq:LRS}
\udot^a=\Omega^a =\Sigma_a={\cal E}_a={\cal H}_a=Q^a=\Pi^a=a_a=0,\\  \Sigma_{ab}={\cal E}_{ab}={\cal H}_{ab}=\Pi_{ab}=\zeta_{ab}=0.
\ea 
Thus the remaining variables are
\ba\label{eq:LRSvar}
{\cal D}_1:&=&\{\udot,\Theta, \Omega, \Sigma, {\cal E}, {\cal H}, \mu, p, Q, \Pi,  \phi, \xi\}\\
&=& {\cal D}_{matter} + {\cal D}_{geometry},
\ea
where 
\be
{\cal D}_{matter}:=\{\mu, p, Q, \Pi\}\,,
\ee
are the matter variables that completely specify the energy momentum tensor of the matter. On the other hand 
\be
{\cal D}_{geometry}:=\{\udot,\Theta, \Omega, \Sigma, {\cal E}, {\cal H}, \phi, \xi\},
\ee
are the geometrical variables.
By decomposing the Ricci identities for $u^a$ and $e^a$ and the doubly contracted Bianchi identities, we then get the following field equations for LRS spacetimes.\\
\medskip \\
\textit{Evolution}:
\ba
   \dot\phi &=& \bra{\sfr23\Theta-\Sigma}\bra{\udot-\sfr12\phi}
+2\xi\Omega+Q\ , \label{phidot}
\\ 
\dot\xi &=& \bra{\sfr12\Sigma-\sfr13\Theta}\xi+\bra{\udot-\sfr12\phi}\Omega
\nonumber \\ && +\sfr12 \hh,  \label{xidot}
\\
\dot\Omega &=& \udot\xi+\Omega\bra{\Sigma-\sfr23\Theta}, \label{dotomega}
\\
\dot \hh &=& -3\xi\E+\bra{\sfr32\Sigma-\Theta}\hh+\Omega Q
\nonumber\\ && +\sfr32\xi\Pi,
\ea
\smallskip

\textit{Propagation}:
\ba
\hat\phi  &=&-\sfr12\phi^2+\bra{\sfr13\Theta+\Sigma}\bra{\sfr23\Theta-\Sigma}
    \nonumber\\&&+2\xi^2-\sfr23\bra{\mu+\Lambda}
    -\E -\sfr12\Pi,\,\label{hatphinl}
\\
\hat\xi &=&-\phi\xi+\bra{\sfr13\Theta+\Sigma}\Omega , \label{xihat}
\\
\hat\Sigma-\sfr23\hat\Theta&=&-\sfr32\phi\Sigma-2\xi\Omega-Q\
,\label{Sigthetahat}
 \\
  \hat\Omega &=& \bra{\udot-\phi}\Omega, \label{Omegahat}
\\
\hat\E-\sfr13\hat\mu+\sfr12\hat\Pi&=&
    -\sfr32\phi\bra{\E+\sfr12\Pi}+3\Omega\hh
 \nonumber\\&&   +\bra{\sfr12\Sigma-\sfr13\Theta}Q , \label{Ehatmupi}
\\
\hat \hh &=& -\bra{3\E+\mu+p-\sfr12\Pi}\Omega
\nonumber\\&&-\sfr32\phi \hh-Q\xi,
\ea
\smallskip

\textit{Propagation/evolution}:
\ba
   \hat\udot-\dot\Theta&=&-\bra{\udot+\phi}\udot+\sfr13\Theta^2
    +\sfr32\Sigma^2 \nonumber\\
    &&-2\Omega^2+\sfr12\bra{\mu+3p-2\Lambda}\ ,\label{Raychaudhuri}
\\
    \dot\mu+\hat Q&=&-\Theta\bra{\mu+p}-\bra{\phi+2\udot}Q \nonumber \\
&&- \sfr32\Sigma\Pi,\,
\\    \label{Qhat}
\dot Q+\hat
p+\hat\Pi&=&-\bra{\sfr32\phi+\udot}\Pi-\bra{\sfr43\Theta+\Sigma} Q\nonumber\\
    &&-\bra{\mu+p}\udot\ ,
\ea 
\ba
\dot\Sigma-\sfr23\hat\udot
&=&
\sfr13\bra{2\udot-\phi}\udot-\bra{\sfr23\Theta+\sfr12\Sigma}\Sigma\nonumber\\
        &&-\sfr23\Omega^2-\E+\sfr12\Pi\, ,\label{Sigthetadot}
\\  
\dot\E +\sfr12\dot\Pi +\sfr13\hat Q&=&
    +\bra{\sfr32\Sigma-\Theta}\E-\sfr12\bra{\mu+p}\Sigma \nonumber \\
  && -\sfr12\bra{\sfr13\Theta+\sfr12\Sigma}\Pi+3\xi\hh \nonumber\\
    &&+\sfr13\bra{\sfr12\phi-2\udot}Q,
\label{edot}
\ea

\textit{Constraint:}
\be
\hh = 3\xi\Sigma-\bra{2\udot-\phi}\Omega. \label{H}
\ee

\section{Most general class of LRS spacetimes}\label{sec:3}

As described in \cite{Elst_Ellis_1996}, if we consider a perfect fluid form of matter with $Q=\Pi=0$, then the propagation equations 
 evolve consistently in time if and only if \begin{equation}
\Omega\xi=0.
\end{equation} 
The above relation then naturally divides perfect fluid LRS spacetimes in three distinct subcalsses \cite{Ellis_1968,Elst_Ellis_1996}: 
\begin{enumerate}
\item LRS class I: ($\Omega\neq 0, \xi=0$) These are stationary inhomogeneous rotating solutions.
\item LRS class II: ($\xi=0=\Omega$) These are inhomogeneous orthogonal family of solutions that can be both static or dynamic. Spherically symmetric solutions are a subclass of this class. 
\item LRS class III ($\xi\neq 0, \Omega= 0$):These are homogeneous orthogonal models with a spatial twist.
\end{enumerate}

In a recent paper \cite{Singh:2016qmr} we established the existence of and found the necessary and sufficient conditions for the  general class of solutions of Locally Rotationally Symmetric spacetimes that have non vanishing rotation and spatial twist simultaneously: that is for this class of spacetimes we 
have by definition 
\be\label{omegaxi}
\Omega\xi\neq0. 
\ee
By the above, these solutions must be imperfect fluid models.  
We also provided a brief algorithm indicating how to solve the system of field equations with the given Cauchy data on an initial spacelike Cauchy surface. The important features of this class of spacetimes are as follows:
\begin{enumerate}
\item  The necessary condition for a LRS spacetime to have non-zero rotation and spatial twist simultaneously is the presence of non-zero heat flux $Q$ which is bounded from both sides.  
\item In these spacetimes {\it all} scalars $\Psi$ obey the following consistency relation:
\be\label{scalarcons}
\forall \Psi, \,\,\, \dot\Psi\Omega = \hat\Psi \xi,
\ee
This equation can be easily derived by noting that for any scalar $\Psi$ in a general LRS spacetime we have $\nabla_a\Psi=-\dot{\Psi}u_a+\hat\Psi e_a$ and $\epsilon^{ab}\nabla_a\nabla_b \Psi=0$.
Also the above equation (\ref{scalarcons}), which is required by (\ref{omegaxi}),  implies self-similarity, for it applies to all scalars, and is unchanged under the transformation $\tau\rightarrow a\tau,$ $\rho \rightarrow a\rho,$ where $\tau$ and $\rho$ are the curve parameters of the integral curves of $u$ and $e$ respectively.
\item The above symmetries generate further constraints and hence the total set of  constraint equations  are now  $\mathcal{C}\equiv\{\mathcal{C}_1,\mathcal{C}_2,\mathcal{C}_3,\mathcal{C}_4\}$, where  
\ba
\mathcal{C}_1:= \hh& = & 3\xi\Sigma-\bra{2\udot+\frac{\Omega}{\xi}\left(\Sigma-\sfr23\Theta\right)}\Omega\,,  \label{constraint0}\\
\mathcal{C}_2:= \phi&=& -\frac{\Omega}{\xi}\left(\Sigma-\sfr23\Theta\right) \,, \label{constraint1} \\
\mathcal{C}_3:=Q&=&-\sfr{\sfr{\Omega}{\xi}}{1+\bra{\sfr{\Omega}{\xi}}^2}\left(\mu+p+\Pi\right),\label{omegaxi2} \\ 
\mathcal{C}_4:=\E &=&  \frac{\Omega}{\xi} \udot \left(\Sigma-\frac23\Theta\right)-\Sigma^2+\frac13\Theta\Sigma +\frac29\Theta^2 \nonumber \\
&&+2\left(\xi^2-\Omega^2\right)+\sfr{\bra{\sfr{\Omega}{\xi}}^2}{1+\bra{\sfr{\Omega}{\xi}}^2}\left(\mu+p+\Pi\right) \nonumber \\
&&-\frac12\Pi-\frac23\mu\;. \label{E}
\ea
\end{enumerate}

It is important to verify that all these new constraints evolve consistently  in time. This is indeed the case, as these constraints are derived by taking all the scalars 
$\Psi\in{\cal D}_1$ and using the equation (\ref{scalarcons}) (which is true for all epochs) together with the field equations. Therefore the time derivatives of these new constraints will identically vanish using (\ref{scalarcons}) and the field equations as we feed the solutions back to the same system.
Therefore solving for the set of variables
\be
{\cal D}_2 :=\brac{\udot, \Theta, \xi, \Sigma, \Omega, \mu
} ,
\ee
will automatically specify the rest 
\be
{\cal D}_3 :=\brac{Q, \phi, \E, \H,p}, 
\ee
{where we assume an 
equation of state for $p$
of the form 
\begin{equation}\label{eos}
p=p(\mu,\Pi,Q).
\end{equation} We note that the anisotropic pressure $\Pi$ is not restricted 
 by the constraints:  there is no algebraic equation 
 linking it to other thermodynamic variables.} Hence this quantity should be specified at any initial Cauchy surface separately (subject to the energy conditions) and it would then evolve in time, via the field equations.

\section{The reduced set of field equations for self similar variables}\label{Sec4}

We will now use the property of self similarity for the most general class of LRS spacetimes to further reduce the set of independent field equations. Let us consider the set of variables
\be
{\cal D}_4 :=\brac{\udot, \Theta, \xi, \Sigma, \Omega}\subset{\cal D}_2 ,
\ee
Then from the kinematical equations for LRS spacetimes
\ba
\nabla_a u_b &=&-u_ae_b\udot+\e_a\e_b\bra{\sfr13\Theta+\Sigma}+\Omega\lc_{ab} \nonumber\\&&+\N_{ab}\bra{\sfr13\Theta-\sfr12\Sigma},\\
{D}_{a}e_{b} &=& \frac{1}{2}\phi N_{ab} + \xi\epsilon_{ab},
\ea
it is clear that for any element $f\in{\cal D}_4$, we must have
\be\label{kinvar}
f(\tau , \rho) = af(a\tau,a\rho),
\ee
as $u^a$, $e^a$, $N^{ab}$ and $\epsilon^{ab}$ are dimensionless. 
Hence without any loss of generality, all these quantities can be written as
\be\label{fform}
f \equiv \frac{f_0(z)}{\rho},
\ee
where 
\be
z = \frac{\tau}{\rho}\,,
\ee
and $f_0$ is dimensionless. Also, from the Einstein field equations $G_{ab}=T_{ab}$, we can easily see, as before, that all elements $g\in{\cal D}_5$, where
\be
{\cal D}_5 :=\brac{
\mu,\Pi}={\cal D}_2-{\cal D}_4\,,
\ee
must satisfy
\be\label{dynvar}
g(\tau, \rho) =a^2g(a\tau , a\rho).\\
\ee
Therefore these quantities can be generally written as
\be\label{gform}
g \equiv \frac{g_0(z)}{\rho^2}.
\ee
Now the {\it dot} and {\it hat} derivatives of all these elements can be written in terms of the dimensionless variable $z$, in the following way: for $f\in{\cal D}_4$, 
\ba
\dot f &=& \frac{f_{0,z}}{\rho^2},\label{fdot} \\
\hat f &=& -\frac{\left(f_0+zf_{0,z}\right)}{\rho^2},\label{fhat} \ea
and for $g\in{\cal D}_5$\ba
\dot g &=& \frac{g_{0,z}}{\rho^3}, \label{gdot} \\
\hat g &=& -\frac{\left(2g_0+zg_{0,z}\right)}{\rho^3}\label{ghat}. 
\ea
Using the above results, the non-trivial field equations become the following ordinary differential equations:
\ba
\phi_{0,z} &=& \left[\sfr23\Theta_0-\Sigma_0\right]\left[\udot_0-\sfr12\phi_0\right] +2\xi_0\Omega_0+Q_0, \\
\xi_{0,z }&=& \left[\sfr12\Sigma_0-\sfr13\Theta_0\right]\xi_0 +\left[\udot_0-\sfr12\phi_0\right]\Omega_0 \nonumber \\ 
&&+ \sfr12\H_0,\label{xi0z1} \\
\Omega_{0,z} &=& \udot_0\xi_0+\Omega_0\left[\Sigma_0-\sfr23\Theta_0\right], \label{omega0z1}\\
\H_{0,z} &=& -3\xi_0\E_0 + \left[\sfr32\Sigma_0-\Theta_0\right]\H_0 + \Omega_0Q_0  \nonumber \\
&&+\sfr32\xi_0\Pi_0.
\ea
\ba
\Sigma_{0,z} -\sfr23\Theta_{0,z} &=&  -\phi_0\udot_0+\sfr29\Theta_0{}^2+\sfr12\Sigma_0{}^2-2\Omega_0{}^2 \nonumber \\
&&+\sfr13\mu_0+p_0-\sfr23\Theta_0\Sigma_0-\E_0 \nonumber \\
&&+\sfr12\Pi_0,  
\ea
\ba
\E_{0,z}+-\sfr13\mu_{0,z}+\sfr12\Pi_{0,z}&=& +\left[\sfr32\Sigma_0-\Theta_0\right]\E_0 +3\xi_0\H_0 \nonumber \\
&&-\sfr13\left(\mu_0+p_0\right)+\sfr12Q_0\phi_0 \nonumber \\
&& -\left(\sfr16\Theta_0-\sfr14\Sigma_0\right)\Pi_0\nonumber \\
&&-\sfr12\left[\mu_0+p_0\right]\Sigma_0.
\ea
It can be shown that the rest of the field equations become redundant when the following set of dimensionless constraints   $\mathcal{\tilde{C}}\equiv\{\mathcal{\tilde{C}}_1,\mathcal{\tilde{C}}_2,\mathcal{\tilde{C}}_3,\mathcal{\tilde{C}}_4\}$ hold, which are easily derived by using equations (\ref{fform}) and (\ref{gform}) on the set of original constraints $\mathcal{C}$:
\ba
\mathcal{\tilde{C}}_1: \H_0 &=& 3\xi_0\Sigma_0-\left[2\udot_0+\frac{\Omega_0}{\xi_0}\left(\Sigma_0-\sfr23\Theta_0\right)\right]\Omega_0, \label{con1}\\
\mathcal{\tilde{C}}_2: \phi_0&=&-\frac{\Omega_0}{\xi_0}\left(\Sigma_0-\sfr23\Theta_0\right),\label{con2}\\
\mathcal{\tilde{C}}_3: Q_0&=&-\frac{\sfr{\Omega_0}{\xi_0}}{1+\bra{\sfr{\Omega_0}{\xi_0}}^2}\left(\mu_0+p_0+\Pi_0\right),\label{con3}\\
\mathcal{\tilde{C}}_4: \E_0 &=& \frac{\Omega_0}{\xi_0}\udot_0(\Sigma_0-\sfr23\Theta_0)-\Sigma_0{}^2+\sfr13\Theta_0\Sigma_0 \nonumber \\
&&+\sfr29\Theta_0{}^2+2(\Sigma_0{}^2-\Omega_0{}^2)\nonumber \\
&&+\sfr{\bra{\sfr{\Omega_0}{\xi_0}}^2}{1+\bra{\sfr{\Omega_0}{\xi_0}}^2}\left(\mu_0+p_0+\Pi_0\right)\nonumber \\&&-\sfr12\Pi_0-\sfr23\mu_0.\label{con4}
\ea

\section{General Solution to the field equations}\label{Sec5}

To find the general solution of the reduced set of the field equations, we note that these equations along with (\ref{scalarcons}), generate the constraint set  $\mathcal{\tilde{C}}\equiv\{\mathcal{\tilde{C}}_1,\mathcal{\tilde{C}}_2,\mathcal{\tilde{C}}_3,\mathcal{\tilde{C}}_4\}$. Hence the field equations are encoded in (\ref{scalarcons}) and the set of constraints, and it suffices to solve (\ref{scalarcons}) along with the constraint to obtain a complete solution to the spacetime. Hence we use equations (\ref{fdot},\ref{fhat},\ref{gdot},\ref{ghat}) in (\ref{scalarcons}) and obtain
\ba
\frac{f_{0,z}}{f_0} &=& \frac{-\xi_0}{\Omega_0+z\xi_0}, \label{eqnf} \\
\frac{g_{0,z}}{g_0} &=& \frac{-2\xi_0}{\Omega_0+z\xi_0}. \label{eqng}
\ea
Now letting $f_0=\Omega_0$ we get 
\be
\frac{\Omega_{0,z}}{\Omega_0}=\frac{-\xi_0}{\Omega_0+z\xi_0},  
\ee
and letting $f_0=\xi_0$ we get 
\be
\frac{\xi_{0,z}}{\xi_0}=\frac{-\xi_0}{\Omega_0+z\xi_0}.  
\ee
The above two equations are coupled first order ordinary differential equations for $\Omega_0$ and $\xi_0$ and the general solution is given by
\ba
\xi_0 (z) &=& -\frac{1}{Az+B}, \\
\Omega_0 (z) &=& -\frac{B}{A(Az+B)}, 
\ea
where $A$ and $B$ are constants of integration. Now using these solutions in (\ref{eqnf},\ref{eqng}), we get the following decoupled equations
\ba
\frac{f_{0,z}}{f_0} &=& -\frac{A}{Az+B}, \label{f_0,z}\\
\frac{g_{0,z}}{g_0} &=& -\frac{2A}{Az+B}. \label{g_0,z}
\ea
The general solutions for the equations (\ref{f_0,z}) and (\ref{g_0,z}) are given by
\ba
f_0 &=& \frac{C_f}{Az+B}, \\
g_0 &=& \frac{C_g}{(Az+B)^2}.
\ea
Here $C_f$ and $C_g$ are integration constants related to each of the kinematic and dynamic variables $f_0$ and $g_0$. Thus the set `$C$' of arbitrary integration constants that we must specify to obtain the general solution for the most general LRS spacetime is given by:
\be
C\equiv\bra{A, B, C_\udot, C_\Theta, C_\Sigma, C_\mu, C_\Pi}\,,
\ee
where we must have $A\ne0$ and $B\ne0$ for the equation (\ref{omegaxi}) to be true. The rest of the variables can then be easily obtained by using the constraint equations.

 For example, using the constraint $\mathcal{\tilde{C}}_1$  (equation (\ref{con1})) we get the magnetic part of the Weyl scalar as follows:
\be
\H=\frac{C_\H}{(Az+B)^2}\;,
\ee
where we have 
\be
C_\H=-3C_\Sigma+\left(2C_\udot+\frac{B}{A}(C_\Sigma-\sfr23C_\Theta)\right)\frac{B}{A}\;.
\ee
Again, using the constraint $\mathcal{\tilde{C}}_2$  (equation (\ref{con2})) we get
\be
\phi_0=\frac{C_\phi}{Az+B}\;\;;\;\; C_\phi=-\frac{B}{A}(C_\Sigma-\sfr23C_\Theta)
\ee
The variables $Q_0$ and $\E_0$ can similarly be obtained using equations (\ref{con3}) and (\ref{con4}) subject to the dimensionless algebraic equation of state $p_0=p_0(\mu_0,Q_0,\Pi_0)$, which must be provided separately along with the field equations. Once an equation of state in form of (\ref{eos}) is given, it is in principle possible to obtain such a dimensionless equation of state, as all the elements of ${\cal D}_{matter}$ have the same symmetries as  (\ref{dynvar}) and hence the dimensionless part can be extracted from all of them. 

Thus we obtain the solution for all the scalar variables of the set ${\cal D}_1$ which completes the general solution. One can in principle obtain the metric elements from the definition of these covariant scalars. However it is important to note that all the physical properties of the LRS spacetime can be obtained directly from these covariant scalars as all of them have well defined geometrical and physical meaning. In the next section we will discuss some of the physical properties of these solutions for both astrophysical and cosmological scenarios. 

\section{Cosmological and Astrophysical properties of this general solution}\label{Sec6}

This class of solutions have some very interesting properties, for both cosmological and stellar collapse scenarios which we list below. We can immediately see that there is a spacetime singularity along the curve $B\rho+A\tau=0$, which is similar to the cosmological singularity of the FLRW or Lemaitre-Tolman-Bondi universes (or corresponding black hole singularities if we take the collapsing branch of the solutions). Apart from this, there are no other singular points on the manifold. 
\begin{enumerate}
\item The most interesting feature of the singularity in this class of spacetime is it can be made timelike, spacelike or null by choice of the ratio of the constants $A$ and $B$. In other words, the ratio of rotation ($\Omega$) and spatial twist ($\xi$) at any initial Cauchy surface completely determines the nature of the initial (or final) singularity and this gives a range of different possibilities.
 \item  For the cosmological scenario, let us consider both $A$ and $B$ to be greater than zero, In that case the initial singularity is along the line $B\rho+A\tau=0$. This `{\it Big Bang}' is no longer instantaneous, and can be spacelike, timelike or null. Thus the section of the manifold that depicts the universe is given by 
\be
\rho>0 \;\;,\;\;\tau> -(B/A)\rho\;.
\ee
For an expanding universe with positive energy density, we must have  $\Theta>0$ and $\mu>0$, and hence we must choose the constants
\be
C_\Theta>0\;\;;\;\; C_\mu>0\;.
\ee
For the cosmological case we can choose {\it dustlike} matter with 
\be
p_0=0\;,
\ee
\be
C_\Pi=0 \Rightarrow \Pi_0=0 .
\ee
Now we can immediately see that in this case $\dot{\Theta}<0, \dot{\mu}<0$. There is no bounce in this cosmology as the expansion goes to zero asymptotically.  Furthermore it is interesting to note that at spacelike infinity `$i_0$' (where $\rho\rightarrow\infty$), timelike infinity `$i+$'  (where $\tau\rightarrow\infty$) and future null infinity $\mathcal{I}+$, all the kinematical and dynamical quantities vanish, making the spacetime asymptotically  Minkowski. Hence, we get a cosmology that is {\it Future asymptotically simple}.
\item Another interesting case happens when the curves $B\rho+A\tau=\rm{const.}$ are null. In this case the initial singularity is {\it incoming null}. Then for any observer on the worldline 
$\rho=0, (\tau>0)$,  observation along the past null cone will depict a universe with homogeneous density, in contrast to the fact that 
on a given time slice the density is inhomogeneous.\\

\item A similar picture can be obtained for collapsing stellar configurations with $A<0$ and $B>0$. In that case the section of the manifold $\rho>0$ and $\tau<(B/|A|)\rho$ depicts the regular collapsing region which is {\it Past asymptotically simple}. To get a collapsing branch of the solution with positive matter density we must have 
$\Theta<0$ and $\mu>0$. Hence we choose
\be
C_\Theta<0\;\;;\;\; C_\mu>0\ .
\ee
Also here we should specify the equation of state linking the  isotropic pressure to other thermodynamic variables and separately specify the constant $C_\Pi$ at the initial Cauchy surface subject to the energy conditions.
We can easily check that in this case $\dot{\Theta}<0, \dot{\mu}>0$. Hence the collapse continues till $\Theta\rightarrow -\infty$ and $\mu\rightarrow\infty$. This is a final singularity at $\tau=(B/|A|)\rho$ and we can easily see that this singularity can be timelike, spacelike, or null, which will have important consequences in terms of the cosmic censorship conjecture.
\end{enumerate}

 \begin{acknowledgments}
SS, GFRE and RG are supported by National Research Foundation (NRF), South Africa. SDM 
acknowledges that this work is based on research supported by the South African Research Chair Initiative of the Department of
Science and Technology and the National Research Foundation.
\end{acknowledgments}

\end{document}